\documentclass[envcountsame,runningheads]{llncs}
\usepackage{amsmath}
\usepackage{setspace}
\usepackage{amssymb}
\usepackage{enumerate}
\usepackage{amsfonts}
\usepackage{graphicx}
\usepackage{xspace}
\usepackage{mathrsfs}
\usepackage{color}
\usepackage{url}
\usepackage{fancyvrb}
\usepackage{wrapfig, framed, caption}
\usepackage{multicol}
\newcounter{instr}

\newcommand{\BDM}{BDM\xspace}

\newcommand{\bigO}{\mathcal{O}}

\def\pp{\mathinner{\,\ldotp\ldotp\,}}

\newcommand{\best}[1]{\underline{\textbf{#1}}}

\begin{document}

\title{Speeding Up String Matching\\by Weak Factor Recognition\thanks{This paper will appear in proceedings of the Prague Stringology Conference 2017}}

\author{Domenico Cantone$^\dag$, Simone Faro$^\dag$, and Arianna Pavone$^\ddag$}
\institute{
$^\dag$Universit\`a di Catania, Viale A. Doria 6, 95125 Catania, Italy\\
$^\ddag$Universit\`a di Messina, Via Concezione 6, 98122 Messina, Italy\\
}

\maketitle

\begin{abstract}
String matching is the problem of finding all the substrings of a text which match a given pattern. It is one of the most investigated problems in computer science, mainly due to its very diverse applications in several fields. 
Recently, much research in the string matching field has focused on the efficiency and flexibility of the searching procedure and quite effective techniques have been proposed for speeding up the existing solutions. In this context, algorithms based on factors recognition are among the best solutions. 

In this paper, we present a simple and very efficient algorithm for string matching based on a weak factor recognition and hashing. Our algorithm has a quadratic worst-case running time. 
However, despite its quadratic complexity, experimental results show that our algorithm obtains in most cases the best running times when compared, under various conditions, against the most effective algorithms present in literature. 
In the case of small alphabets and long patterns, the gain in running times reaches 28\%. This makes our proposed algorithm one of the most flexible solutions in practical cases.\\


\end{abstract}

\section{Introduction} \label{sec:introduction}
The \emph{exact string matching problem} is one of the most studied problem in computer science. It consists in finding all the (possibly overlapping) occurrences of an input pattern $x$ in a text $y$, over a given alphabet $\Sigma$ of size $\sigma$. A huge number of solutions has been devised since the 1980s \cite{charras04,FL13} and, despite such a wide literature, much work has been produced in the last few years, indicating that the need for efficient solutions to this problem is still high.

Solutions to the exact string matching problem can be divided in two classes: \emph{counting} solutions simply return the number of occurrences of the pattern in the text, whereas \emph{reporting} solutions provide also the exact positions at which the pattern occurs. Solutions in the first class are in general faster than the ones in the second class. In this paper we are interested in algorithms belonging to the class of reporting solutions.

From a theoretical point of view, the exact string matching problem has been studied extensively. If we denote by $m$ and $n$ the lengths of the pattern and of the text, respectively, the problem can be solved in $\bigO(n)$ worst-case time complexity \cite{KMP77}. However, in many practical cases it is possible to avoid reading all the characters of the text, thus achieving sublinear performances on the average. The optimal average $\bigO(\frac{n\log_{\sigma}m}{m})$ time complexity \cite{yao79} has been reached for the first time by the Backward DAWG Matching algorithm \cite{crochemore94} (BDM).  
However, all algorithms with a sublinear average behaviour may have to possibly read all the text characters in the worst case. It is interesting to note that many of those algorithms have an $\bigO(nm)$-time complexity in the worst-case. Interested readers can refer to \cite{charras04,FL10,FL13} for a detailed survey of the most efficient solutions to the problem.


The BDM algorithm computes the Directed Acyclic Word Graph (DAWG) of the reverse $x^{\mathsf{R}}$ of the pattern $x$. Such graph is an automaton which recognizes all and only the factors of $x^{\mathsf{R}}$, and can be computed in $\bigO(m)$ time. 
During the searching phase, the BDM algorithm moves a window of size $m$ on the text. For each new position of the window, the automaton of $x^{\mathsf{R}}$ is used to search for a factor of $x$ from the right to the left of the window. The basic idea of the BDM algorithm is that when the backward search fails on a letter $c$ after reading a word $u$, then $cu$ can not be a factor of $p$, so that moving the window just after $c$ is safe.
In addition, the algorithm maintains the length of the last recognized suffix of $x^\mathsf{R}$, which is a prefix of the pattern. If a suffix of length $m$ is recognized, then an occurrence of the pattern is reported. 

We say that the DAWG of a string performs an \emph{exact factor recognition} since the accepted language coincides exactly with the set of the factors of the string. On the other hand, we say that a structure performs a \emph{weak factor recognition} when it is able to recognize \emph{at least} all the factors of the string, but maybe something more.
For instance, the Factor Oracle \cite{ACR99} of a string $x$ performs a weak factor recognition of the factors of $x$. It is an automaton which recognizes all the factors of $x$ acting like an oracle: if a string is accepted by the automaton, it \emph{may be} a factor of $x$. However, all the factors of $x$ are accepted.
Due to its relaxed recognition approach, the Factor Oracle can be constructed and handled using less resources than the DAWG, both in terms of space and time.  

The Backward Oracle Matching algorithm \cite{ACR99} (BOM) works in the same way as the BDM algorithm, but makes use of the Factor Oracle of the reverse pattern, in place of the DAWG. In practical cases, the resulting algorithm performs better than the BDM algorithm \cite{FL13}.

Both BDM and BOM algorithms have been recently improved in various way. For instance, very fast \BDM-like algorithms based on the bit-parallel simulation of the nondeterministic factor automaton~\cite{BYG92} have been presented in~\cite{NR98}, whereas efficient extensions of the BOM algorithm appeared in \cite{FL09}.


In this paper we present a new fast string matching algorithm based on a(n) (even more) weak factor recognition approach. Our solution uses a hash function to recognize all the factors of the input pattern. Such method leads to a simple and very fast recognition mechanism and makes the algorithm very effective in practical cases. 
In Section \ref{sec:new}, we introduce and analyze our proposed algorithm, whereas in Section \ref{sec:results} we compare experimentally its performance against the most effective solutions present in the literature. Finally, we draw our conclusions in Section \ref{sec:conclusions}.

\section{An Efficient Weak-Factor-Recognition Approach} \label{sec:new}
In this section we present an efficient algorithm for the exact string matching problem based on a weak-factor-recognition approach with hashing.  
Though the resulting algorithm has a quadratic worst-case time complexity, on average it shows a sublinear behaviour.

Let $x$ be a pattern of length $m$ and $y$ a text of length $n$. In addition, let us assume that both strings $x$ and $y$ are drawn from a common alphabet $\Sigma$ of size $\sigma$.
Our proposed algorithm, named \emph{Weak Factor Recognition} (\textsc{Wfr}) is able to count and report all the occurrences of $x$ in $y$.
It consists in a preprocessing and a searching phase. These are described in detail in the following sections.

\subsection{The Preprocessing Phase}

During the preprocessing phase, all subsequences of the pattern $x$ are indexed to facilitate their search during the searching phase. 
Specifically, we define a hash function $h: \Sigma^* \rightarrow \{0 \pp 2^{\alpha}-1\}$, which associates an integer value $0 \leq v < 2^{\alpha}$ (for a given bound $\alpha$)\footnote{In our setting, the value $\alpha$ has been fixed to ${16}$, so that each hash value fits into a single $16$-bit register.} with any string over the alphabet $\Sigma$.
Here, we shall make the assumption that each character $c\in \Sigma$ can be handled as an integer value, so that arithmetic operations can be performed on characters. For instance, in many practical applications, input strings can be handled as sequences of ASCII characters. Thus each character can be seen as an 8-bit value corresponding to its ASCII code.

For each string $x\in \Sigma^*$ of length $m\geq 0$, the value of $h(x)$ is recursively defined as follows
$$
	h(x) := \left\{
		\begin{array}{ll}
			0 & \textrm{ if } m = 0\\
			\left( h(x[1 \pp m-1]) \times 2 + x[0] \right) \mod 2^{\alpha}  ~~~~& \textrm{ otherwise}.
		\end{array}
		\right.
$$
Observe that, for each string $x\in \Sigma^{*}$, we have $0\leq h(x) < 2^{\alpha}$.

The preprocessing phase of our algorithm, which is reported in Fig.~\ref{fig:codesks} (on the left), consists in
computing the hash values of all possible substrings of the pattern $x$.  

A bit vector $F$ of size $2^{\alpha}$ is maintained for storing the hash values corresponding to the factors of $x$. Thus, if $z$ is a factor of $x$, then the bit at position $h(z)$ in $F$ is set (i.e., $F[h(z)] :=1$), otherwise it is set to $0$.
More formally, for each value $v$ in the bit vector, with $0\leq v < 2^{\alpha}$, we have
$$
	F[v] := \left\{
		\begin{array}{ll}
			1 & \text{if } h(x[i \pp j])=v \text{, for some } 0\leq i \leq j < m \\
			0 ~~~~~~& \text{otherwise}.
		\end{array}
		\right.
$$
Given two strings $x,z\in \Sigma^{*}$, it is easy to prove that if $z$ is a factor of $x$ then $F[h(z)]=1$; on the other hand, when $F[h(z)]=1$, in general we can not conclude that $z$ is a factor of $x$.


Let $w$ be the number of bits in a computer word of the target machine. Then the bit vector $F$ can be implemented as a table of $2^{\alpha}/w$ words.\footnote{In our setting, we have $w=8$ and $F$ has been implemented as a table of $8,192$ chars, corresponding to a bit-vector of 65,536 bits.}
The procedure \textsc{SetBit($F,i$)} and the function \textsc{TestBit($F,i$)} (both reported in Fig.~\ref{fig:codesks}, on the left) are used to quickly set and query, respectively, the bit at position $i$ in the vector $F$. Such procedures are very fast and can be executed in constant time.

Since the set of all nonempty factors of a string $x$ of length $m$ has size $m^2$, the preprocessing phase of the algorithm requires $\bigO(2^{\alpha})$ space and $\bigO(m^2)$ time.

\begin{figure}[!t]
\begin{center}
\begin{tabular}{|ll|}
\hline
\begin{tabular}{rl}
&\\
\multicolumn{2}{l}{~~\textsc{SetBit($F, v$)}}\\
~\textsf{1.} & \textsf{$p \leftarrow \lfloor v/w \rfloor$}\\ 
~\textsf{2.} & \textsf{$b \leftarrow v$ mod $w$}\\ 
~\textsf{3.} & \textsf{$F[p] \leftarrow F[p]$ or $(1 \ll b)$}\\ 
&\\
\multicolumn{2}{l}{~~\textsc{TestBit($F, v$)}}\\
~\textsf{1.} & \textsf{$p \leftarrow \lfloor v/w \rfloor$}\\ 
~\textsf{2.} & \textsf{$b \leftarrow v$ mod $w$}\\ 
~\textsf{3.} & \textsf{return $(F[p]$ and $(1 \ll b)) \neq 0$}\\ 
&\\
\multicolumn{2}{l}{~~\textsc{Preprocessing($x, m$)}}\\
~\textsf{1.} & \textsf{for $v \leftarrow 0$ to $2^{\alpha}-1$ do}\\
~\textsf{2.} & \quad \textsf{$F[v] \leftarrow$ 0}\\ 
~\textsf{3.} & \textsf{for $i \leftarrow m-1$ downto $0$ do}\\ 
~\textsf{4.} & \quad \textsf{$v \leftarrow 0$}\\ 
~\textsf{5.} & \quad \textsf{for $j \leftarrow i$ downto $0$ do}\\ 
~\textsf{6.} & \quad \quad \textsf{$v \leftarrow (v \ll 2) + x[j]$}\\ 
~\textsf{7.} & \quad \quad \textsc{SetBit($F,v$)}\\ 
~\textsf{8.} & \textsf{return $F$}\\ \\
\end{tabular} &
\begin{tabular}{rl}
\multicolumn{2}{l}{~~\textsc{Check($x, m, y, i$)}}\\
~\textsf{1.} & \textsf{$k \leftarrow 0$}\\ 
~\textsf{2.} & \textsf{while ($k < m$ and $x[k]=y[i+k]$) do}\\ 
~\textsf{3.} & \quad \textsf{$k \leftarrow k+1$}\\ 
~\textsf{4.} & \textsf{if $k = m$ then return true}\\ 
~\textsf{5.} & \textsf{return false}\\ 
&\\
\multicolumn{2}{l}{~~\textsc{Wfr($x, m, y, n,$)}}\\
~\textsf{1.} & \textsc{$F \leftarrow $Preprocessing$(x,m)$}\\ 
~\textsf{2.} & \textsf{$j \leftarrow m-1$}~~~\\ 
~\textsf{3.} & \textsf{while ($j < n$)  do}~~~\\ 
~\textsf{4.} & \quad \textsf{$v \leftarrow y[j]$}\\ 
~\textsf{5.} & \quad \textsf{$i \leftarrow j-m+1$}\\ 
~\textsf{6.} & \quad \textsf{while ($j> i$ and \textsc{TestBit($F,v$)})  do}~~~~\\ 
~\textsf{7.} & \quad \quad \textsf{$j \leftarrow j-1$}\\ 
~\textsf{8.} & \quad \quad \textsf{$v \leftarrow (v \ll 2) + y[j]$}\\ 
~\textsf{9.} & \quad \textsf{if ($j=i$ and \textsc{TestBit($F,v$)}) then}\\ 
~\textsf{10.} & \quad \quad \textsf{if \textsc{Check($x,m,y,i$)} then return $i$}\\ 
~\textsf{11.} & \quad \textsf{$j \leftarrow j+m$}\\ 
\end{tabular}\\ 
\hline
\end{tabular}
\caption{The pseudo-code of the \textsc{Wfr} algorithm and of some auxiliary procedures.}
\label{fig:codesks}
\end{center}
\end{figure}

\subsection{The Searching Phase}
As in the BDM and BOM algorithms, during the searching phase a window of size $m$ is opened on the text, starting at position $0$. After each attempt, the window is shifted to the right until the end of the text is reached. During an attempt at a given position $i$ of the text, the current window is opened on the substring $y[i \pp j]$ of the text, with $j = i+m-1$. Our algorithm starts computing the hash value $h(y[j])$ corresponding to the rightmost character of the window. If the corresponding bit in $F$ is set, then such substring may be a factor of $x$.  In this case, the algorithm computes the hash value of the subsequent substring, namely, $h(y[j-1 \pp j])$.

More precisely, the hash value $y[j-k \pp j]$ of the suffixes of the window is computed for increasing values of $k$, until $k$ reaches the value $m$ or until the corresponding bit in $F$ is not set.

Observe that by using the following relation
$$
	h(y[j-k \pp j]) = \Big( \left(h(y[j-k+1 \pp j]) \ll 1\right) + y[j-k]\Big) \mod 2^{\alpha}\,,
$$
the hash value of the suffix $y[j-k \pp j]$ can be computed in constant time in terms of $h(y[j-k+1 \pp j])$.

When an attempt ends up with $k=m$, a naive check is performed in  order to verify whether the substring $y[i \pp j]$ matches the pattern (see procedure \textsc{Check} shown in Fig.~\ref{fig:codesks}). Such verification can obviously be performed in $\bigO(m)$ time. In this case, the shift advancement is of a single character to the right.

Table \ref{exp_false_positives} shows the average number of occurrences ($\alpha$ value) versus the average number of verifications ($\beta$ value) for every 1024Kb. Values have been computed during the searching phase in our experimental tests described in Section \ref{sec:results}. Notice that the number of exceeding verifications is negligible and, in most cases, equal to 0.

The pseudo-code provided in Fig.~\ref{fig:codesks} (on the right) reports the skeleton of the algorithm. If a naive check were performed after each attempt of the algorithm, then a shift of one position would be performed at each iteration. This leads to a $\bigO(nm)$ worst-case time complexity. However, the experimental results reported in Section~\ref{sec:results} show that, in practical cases, the \textsc{Wfr} algorithm has a sublinear behaviour.

\begin{table}[!t]
\begin{center}
\begin{scriptsize}
\begin{tabular*}{0.99\textwidth}{@{\extracolsep{\fill}}|l| lllllllll |}
\hline
&&&&&&&&&\\[-0.2cm]
~$m$ & $4$ & $8$ & $16$ & $32$ & $64$ & $128$ & $256$ & $512$ & $1024$~~~ \\
&&&&&&&&&\\[-0.2cm]
\hline
&&&&&&&&&\\[-0.1cm]
~Genome-$\alpha$~ & 4068,40	& 23,20 & 0,20	& 0,20 & 0,20 & 0,20 & 0,20 & 0,20 & 0,20\\
~Genome-$\beta$~~  & 4068,40 & 24,40 & 0,20 & 0,20 & 0,20 & 0,20 & 0,20 & 0,20 & 0,20\\
&&&&&&&&&\\
~Protein-$\alpha$~~  & 17,00 & 0,20 & 0,20 & 0,20 & 0,20 & 0,20 & 0,20 & 0,20 & 0,20\\
~Protein-$\beta$~~  & 21,40 & 0,20 & 0,20 & 0,20 & 0,20 & 0,20 & 0,20 & 0,20 & 0,20\\
&&&&&&&&&\\
~English-$\alpha$~~  & 1275,80 & 28,60 & 2,00 & 0,40 & 0,20 & 0,20 & 0,20 & 0,20 & 0,20\\
~English-$\beta$~ & 1280,40 & 28,80 & 2,20 & 0,40 & 0,20 & 0,20 & 0,20 & 0,20 & 0,20\\
&&&&&&&&&\\
\hline
\end{tabular*}
\end{scriptsize}
\end{center}
\caption{\label{exp_false_positives}The average number of occurrences ($\alpha$ value) versus the average number of verifications ($\beta$ value) for every 1024Kb. Values have been computed in the searching phase of the experimental tests described in Section \ref{sec:results}.}
\end{table}

\subsection{Some Improvements}
Practical improvements of the \textsc{Wfr} algorithm can be obtained by means of a \emph{chained-loop} on the characters of the pattern in the implementation of the searching phase.
Such a technique consists in dropping the call to \textsc{TestBit} in the \textsf{while}-loop at line 6, while computing the hash value. The test is performed only every $k$ cycles, for a fixed value of $k$. This leads to a fast computation of the hash values even if the corresponding shifts are shorter on average.

For instance, if $k$ is set to $2$, then lines 4, 7, and 8 of the \textsc{Wfr} algorithm are implemented in the following way:\\

\begin{tabular}{ll}
~~\textsf{4.} & \quad \textsf{$v \leftarrow (y[j] \ll 1) + y[j-1]$}\\ 
~~& \ldots \\ 
~~\textsf{7.} & \quad \quad \textsf{$j \leftarrow j-2$}\\ 
~~\textsf{8.} & \quad \quad \textsf{$v \leftarrow (v \ll 4) + (y[j] \ll 2) + y[j-1]$}\\ 
\end{tabular}\\

The resulting algorithm maintains the same space and time complexity, but in practice it shows a sensible increase of its performance, as shown in the next section.

\section{Experimental Results}\label{sec:results}
We report the experimental results of the performance evaluation of the \textsc{Wfr} algorithm and its variants with a $k$-chained-loop against the most efficient solutions present in literature for the online exact string matching problem.
Specifically, the following 15 algorithms (implemented in 79 variants, depending on the values of their parameters) have been compared:
\begin{itemize}
\item \textsc{AOSO$q$}: the Average-Optimal variant \cite{FG05} of the Shift-Or algorithm \cite{BYG92} using $q$-grams, with $1\leq q \leq 6$;
\item \textsc{BNDM$q$}: the Backward-Nondeterministic-DAWG-Matching algorithm~\cite{NR98} implemented using $q$-grams, with  $1\leq q \leq 8$;
\item \textsc{BSDM$q$}: the Backward-SNR-DAWG-Matching algorithm \cite{FL12} using condensed alphabets with groups of $q$ characters, with $1\leq q \leq 8$;
\item \textsc{BXS$q$}: the Backward-Nondeterministic-DAWG-Matching algorithm~\cite{NR98} with Extended Shift~\cite{DPST10} implemented using $q$-grams, with $1\leq q \leq 8$;
\item \textsc{EBOM}: the extended version~\cite{FL09} of the BOM algorithm~\cite{ACR99};
\item \textsc{FSBNDM$qs$}: the Forward Simplified version~\cite{PT11,FL09} of the BNDM algorithm~\cite{NR98} implemented using $q$-grams $s$-forward characters (with $1\leq q \leq 8$ and $1\leq s\leq 6$);
\item \textsc{KBNDM}: the Factorized variant~\cite{CFG12} BNDM algorithm~\cite{NR98};
\item \textsc{SBNDM$q$}: the Simplified version of the Backward-Nondeterministic-DAWG-Matching algorithm~\cite{ACR99} implemented using $q$-grams, with $1\leq q \leq 8$;
\item \textsc{FS-$w$}: the Multiple Windows version~\cite{FL12b} of the Fast Search algorithm~\cite{CF05} implemented using $w$ sliding windows, with $2\leq w \leq 6$;
\item \textsc{HASH$q$}: the Hashing algorithm~\cite{Lec07} using $q$-grams, with $3\leq q \leq 5$;
\item \textsc{IOM}: the Improved Occurrence Matcher \cite{CF14} 
\item \textsc{WOM}: the Worst Occurrence Matcher \cite{CF14};
\item \textsc{JOM}: the Jumping Occurrence Matcher \cite{CF14};
\item \textsc{WFR}: the new Weak Factors Recognition algorithm;
\item \textsc{WFR$q$}: the new Weak Factors Recognition variants implemented with a $k$-chained-loop (with $2\leq k \leq 4$);
\end{itemize}
For the sake of completeness, we evaluated also the following two string matching algorithms for \emph{counting} occurrences (however, we did not take them into account in our comparison since they simply count the number of matching occurrences):
\begin{itemize}
\item \textsc{EPSM}: the Exact Packed String Matching algorithm \cite{FK14};
\item \textsc{TSO$q$}: the  Two-Way variant of \cite{Durian14} the Shift-Or algorithm \cite{BYG92} implemented with a loop unrolling of $q$ characters, with $q=5$;
\end{itemize}

All algorithms have been implemented in the \textsf{C} programming language and have been tested using the \textsc{Smart} tool \cite{FTBDM16}.\footnote{The \textsc{Smart} tool is available online at  \url{http://www.dmi.unict.it/~faro/smart/}.} All experiments have been executed locally on a MacBook Pro with 4 Cores, a 2 GHz Intel Core i7 processor, 16 GB RAM 1600 MHz DDR3, 256 KB of L2 Cache and 6 MB of Cache L3.
All algorithms have been compared in terms of their running times, including any preprocessing time.

We report experimental evaluations on three real data sets (see Tables \ref{exp_tab_genome}, \ref{exp_tab_protein}, and \ref{exp_tab_english}). Specifically, we used a genome sequence, a protein sequence, and an English text. All sequences have a length of 5MB; they are provided by the\textsc{Smart} research tool and are available online for download.  

In the experimental evaluation, patterns of length $m$ were randomly extracted from the sequences, with $m$ ranging over the set of values $\{2^i \mid 2\leq i \leq 10\}$.
For each case, the mean over the running times (expressed in hundredths of seconds) of $500$ runs has been reported.

The following tables summarize the running times of our evaluations. Each table is divided into four blocks. The first and the second block present the most effective algorithms known in literature based on automata and comparison of characters, respectively. The best results among these two sets of algorithms have been bold-faced in order to easily locate them. The third block contains the running times of our newly proposed algorithm and its variant, including the speed up (in percentage) obtained against the best running time in the first two blocks. Positive values indicate a breaking of the running time whereas a negative percentage represent a performance improvement. Running times which represent an improvement of the performance have been bold-faced.

The last block reports the running times obtained by the best two algorithms for \emph{counting} occurrences (however, as already remarked, these have not been included in our comparison).

\smallskip


\begin{table}
\begin{center}
\begin{scriptsize}
\begin{tabular*}{0.99\textwidth}{@{\extracolsep{\fill}}|l| lllllllll |}
\hline
&&&&&&&&&\\[-0.2cm]
~$m$ & $4$ & $8$ & $16$ & $32$ & $64$ & $128$ & $256$ & $512$ & $1024$ \\
&&&&&&&&&\\[-0.2cm]
\hline
&&&&&&&&&\\[-0.1cm]
~\textsc{AOSO$q$} & 16.98$^{(2)}$ & 9.63$^{(2)}$ & 3.93$^{(4)}$ & 3.39$^{(4)}$ & 2.98$^{(6)}$ & 2.97$^{(6)}$ & 2.99$^{(6)}$ & 3.00$^{(6)}$ & 3.03$^{(6)}$\\
~\textsc{BNDM$q$} & 11.13$^{(4)}$ & 4.10$^{(4)}$ & 2.99$^{(4)}$ & 2.47$^{(4)}$ & 2.38$^{(4)}$ & 2.39$^{(4)}$ & 2.41$^{(4)}$ & 2.47$^{(4)}$ & 2.45$^{(4)}$\\
~\textsc{BSDM$q$} & 8.37$^{(4)}$ & \best{3.71$^{(4)}$} & \best{2.78$^{(4)}$} & \best{2.46$^{(4)}$} & \best{2.25$^{(8)}$} & \best{2.15$^{(8)}$} & \best{2.11$^{(8)}$} & \best{2.16$^{(6)}$} & \best{2.11$^{(6)}$}\\
~\textsc{BXS$q$} & 11.86$^{(2)}$ & 4.78$^{(4)}$ & 3.25$^{(4)}$ & 2.53$^{(6)}$ & 2.50$^{(6)}$ & 2.52$^{(4)}$ & 2.49$^{(4)}$ & 2.55$^{(4)}$ & 2.54$^{(4)}$\\
~\textsc{EBOM} & 7.72 & 7.15 & 5.66 & 4.10 & 3.17 & 2.67 & 2.40 & 2.32 & 2.41\\
~\textsc{FSBNDM$qs$~} & \best{6.46$^{(3,1)}$} & 3.87$^{(4,1)}$ & 2.94$^{(4,1)}$ & 2.38$^{(4,1)}$ & 2.35$^{(6,2)}$ & 2.31$^{(6,1)}$ & 2.33$^{(6,1)}$ & 2.38$^{(3,1)}$ & 2.37$^{(6,1)}$\\
~\textsc{KBNDM} & 10.88 & 8.21 & 6.15 & 4.17 & 3.27 & 3.09 & 3.10 & 3.13 & 3.14\\
~\textsc{SBNDM$q$} & 8.75$^{(2)}$ & 3.95$^{(4)}$ & 2.97$^{(4)}$ & 2.47$^{(4)}$ & 2.39$^{(4)}$ & 2.39$^{(4)}$ & 2.36$^{(4)}$ & 2.38$^{(4)}$ & 2.38$^{(4)}$\\
&&&&&&&&&\\[-0.1cm]
\hline
&&&&&&&&&\\[-0.1cm]
~\textsc{FS-$w$} & 12.33$^{(2)}$ & 9.39$^{(2)}$ & 7.76$^{(2)}$ & 6.89$^{(2)}$ & 6.16$^{(2)}$ & 5.63$^{(2)}$ & 5.06$^{(2)}$ & 4.73$^{(2)}$ & 4.42$^{(2)}$\\
~\textsc{FJS} & 18.60 & 16.69 & 16.96 & 15.96 & 16.09 & 16.80 & 16.71 & 16.61 & 16.59\\
~\textsc{HASH$q$} & 18.09$^{(3)}$ & 7.68$^{(3)}$ & 4.67$^{(5)}$ & 3.31$^{(5)}$ & 2.78$^{(5)}$ & 2.60$^{(5)}$ & 2.63$^{(5)}$ & 2.51$^{(5)}$ & 2.40$^{(5)}$\\
~\textsc{IOM} & 14.41 & 11.88 & 11.08 & 11.17 & 11.17 & 11.13 & 11.03 & 11.03 & 10.98\\
~\textsc{WOM} & 16.69 & 12.48 & 9.88 & 8.61 & 7.75 & 7.16 & 6.72 & 6.29 & 6.11\\
&&&&&&&&&\\[-0.1cm]
\hline
&&&&&&&&&\\[-0.1cm]
~\textsc{WFR} & 13.85 & 8.77 & 5.70 & 3.73 & 2.69 & 2.28 & 1.98 & 1.72 & 1.57\\
~\textsc{WFR$q$}   & 8.67$^{(2)}$ & 4.42$^{(4)}$ & 2.98$^{(4)}$ & \best{2.36$^{(4)}$} & \best{2.08$^{(4)}$} & \best{1.97$^{(4)}$} & \best{1.86$^{(4)}$} & \best{1.62$^{(4)}$} & \best{1.52$^{(4)}$}\\[.2cm]
~\emph{speed-up} & +34\% & +19\% & +7.1\% & -4.0\% & -7.5\% & -8.3\% & -11\% & -25\% & -28\%\\
&&&&&&&&&\\[-0.1cm]
\hline
&&&&&&&&&\\[-0.1cm]
~\textsc{EPSM} & 5.87 & 3.72 & 2.50 & 1.93 & 1.75 & 1.72 & 1.66 & 1.62 & 1.65\\
~\textsc{TSO$q$} & 5.54$^{(5)}$ & 3.85$^{(5)}$ & 3.08$^{(5)}$ & 2.42$^{(5)}$ & 2.05$^{(5)}$ & - & - & - & -\\
&&&&&&&&&\\
\hline
\end{tabular*}
\end{scriptsize}
\end{center}
\caption{\label{exp_tab_genome}Experimental results on a genome sequence.}
\end{table}

\begin{table}
\begin{center}
\begin{scriptsize}
\begin{tabular*}{0.99\textwidth}{@{\extracolsep{\fill}}|l| lllllllll |}
\hline
&&&&&&&&&\\[-0.2cm]
~$m$ & $4$ & $8$ & $16$ & $32$ & $64$ & $128$ & $256$ & $512$ & $1024$ \\
&&&&&&&&&\\[-0.2cm]
\hline
&&&&&&&&&\\[-0.1cm]
~\textsc{AOSO$q$} & 10.80$^{(2)}$ & 4.27$^{(4)}$ & 3.84$^{(4)}$ & 3.81$^{(4)}$ & 3.18$^{(4)}$ & 3.17$^{(4)}$ & 3.16$^{(4)}$ & 3.16$^{(4)}$ & 3.16$^{(4)}$\\
~\textsc{BNDM$q$} & 12.20$^{(4)}$ & 4.29$^{(4)}$ & 3.06$^{(4)}$ & 2.46$^{(4)}$ & 2.45$^{(4)}$ & 2.43$^{(4)}$ & 2.42$^{(4)}$ & 2.40$^{(4)}$ & 2.40$^{(4)}$\\
~\textsc{BSDM$q$} & 4.68$^{(2)}$ & 3.71$^{(2)}$ & 2.75$^{(4)}$ & 2.35$^{(4)}$ & \best{2.06$^{(4)}$} & \best{1.98$^{(4)}$} & \best{1.97$^{(4)}$} & \best{1.97$^{(4)}$} & \best{1.94$^{(4)}$}\\
~\textsc{BXS$q$} & 6.91$^{(2)}$ & 4.29$^{(2)}$ & 3.12$^{(2)}$ & 2.52$^{(2)}$ & 2.48$^{(2)}$ & 2.52$^{(2)}$ & 2.50$^{(2)}$ & 2.51$^{(2)}$ & 2.52$^{(2)}$\\
~\textsc{EBOM} & \best{3.87} & \best{2.94} & \best{2.57} & \best{2.29} & 2.11 & 2.18 & 2.20 & 2.24 & 2.42\\
~\textsc{FSBNDM$qs$~} & 4.32$^{(2,0)}$ & 3.28$^{(2,0)}$ & 2.59$^{(3,1)}$ & 2.26$^{(3,1)}$ & 2.22$^{(3,1)}$ & 2.25$^{(3,1)}$ & 2.25$^{(3,1)}$ & 2.20$^{(3,1)}$ & 2.26$^{(3,1)}$\\
~\textsc{KBNDM} & 7.46 & 4.97 & 3.81 & 3.24 & 3.04 & 3.01 & 2.95 & 2.96 & 2.95\\
~\textsc{SBNDM$q$} & 5.25$^{(2)}$ & 3.67$^{(2)}$ & 2.79$^{(2)}$ & 2.34$^{(2)}$ & 2.45$^{(4)}$ & 2.41$^{(4)}$ & 2.42$^{(4)}$ & 2.41$^{(4)}$ & 2.40$^{(4)}$\\
&&&&&&&&&\\[-0.1cm]
\hline
&&&&&&&&&\\[-0.1cm]
~\textsc{FS-$w$} & 6.18$^{(2)}$ & 4.33$^{(2)}$ & 3.55$^{(2)}$ & 3.20$^{(2)}$ & 3.05$^{(2)}$ & 2.94$^{(2)}$ & 2.90$^{(2)}$ & 2.87$^{(2)}$ & 2.86$^{(2)}$\\
~\textsc{FJS} & 9.68 & 18.54 & 4.18 & 3.02 & 2.92 & 2.89 & 2.82 & 3.16 & 4.11\\
~\textsc{HASH$q$} & 19.92$^{(3)}$ & 8.36$^{(3)}$ & 5.05$^{(3)}$ & 3.75$^{(5)}$ & 3.19$^{(5)}$ & 2.99$^{(5)}$ & 2.92$^{(5)}$ & 2.76$^{(5)}$ & 2.66$^{(5)}$\\
~\textsc{IOM} & 8.87 & 6.36 & 5.02 & 4.41 & 4.04 & 3.92 & 3.86 & 3.86 & 3.79\\
~\textsc{WOM} & 9.31 & 6.61 & 5.13 & 4.32 & 4.03 & 3.72 & 3.56 & 3.43 & 3.33\\
&&&&&&&&&\\[-0.1cm]
\hline
&&&&&&&&&\\[-0.1cm]
~\textsc{WFR}   & 6.79 & 5.80 & 4.43 & 3.21 & 2.65 & 2.38 & 2.12 & 1.87 & 1.70\\
~\textsc{WFR$q$} & 4.85$^{(2)}$ & 3.69$^{(2)}$ & 2.98$^{(4)}$ & 2.36$^{(4)}$ & \best{2.03$^{(4)}$} & \best{1.93$^{(4)}$} & \best{1.89$^{(4)}$} & \best{1.75$^{(4)}$} & \best{1.66$^{(4)}$}\\
~\emph{speed-up} & +25\% & +25\% & +15\% & +3.0\% & -1.4\% & -2.5\% & -4.0\% & -11\% & -14\%\\
&&&&&&&&&\\[-0.1cm]
\hline
&&&&&&&&&\\[-0.1cm]
~\textsc{EPSM} & 6.67 & 5.55 & 2.77 & 2.16 & 1.91 & 1.91 & 1.90 & 1.83 & 1.86\\
~\textsc{TSO$q$} & 5.41$^{(5)}$ & 3.90$^{(5)}$ & 3.29$^{(5)}$ & 2.59$^{(5)}$ & 2.17$^{(5)}$ & - & - & - & -\\
&&&&&&&&&\\
\hline
\end{tabular*}
\end{scriptsize}
\end{center}
\caption{\label{exp_tab_protein}Experimental results on a protein sequence.}
\end{table}

\begin{table}
\begin{center}
\begin{scriptsize}
\begin{tabular*}{0.99\textwidth}{@{\extracolsep{\fill}}|l| lllllllll |}
\hline
&&&&&&&&&\\[-0.2cm]
~$m$ & $4$ & $8$ & $16$ & $32$ & $64$ & $128$ & $256$ & $512$ & $1024$ \\
&&&&&&&&&\\[-0.2cm]
\hline
&&&&&&&&&\\[-0.1cm]
~\textsc{AOSO$q$} & 11.14$^{(2)}$ & 4.58$^{(4)}$ & 3.89$^{(4)}$ & 3.76$^{(4)}$ & 3.16$^{(6)}$ & 3.16$^{(6)}$ & 3.18$^{(6)}$ & 3.21$^{(6)}$ & 3.16$^{(6)}$\\
~\textsc{BNDM$q$} & 12.30$^{(4)}$ & 4.35$^{(4)}$ & 3.17$^{(4)}$ & 2.49$^{(4)}$ & 2.53$^{(4)}$ & 2.52$^{(4)}$ & 2.51$^{(4)}$ & 2.54$^{(4)}$ & 2.51$^{(4)}$\\
~\textsc{BSDM$q$} & 4.73$^{(2)}$ & 3.85$^{(2)}$ & \best{2.86$^{(4)}$} & \best{2.35$^{(4)}$} & \best{2.20$^{(4)}$} & \best{2.09$^{(4)}$} & \best{2.07$^{(4)}$} & \best{2.02$^{(4)}$} & \best{2.00$^{(4)}$}\\
~\textsc{BXS$q$} & 7.38$^{(2)}$ & 4.85$^{(2)}$ & 3.43$^{(4)}$ & 2.59$^{(4)}$ & 2.59$^{(4)}$ & 2.64$^{(4)}$ & 2.62$^{(4)}$ & 2.62$^{(4)}$ & 2.63$^{(4)}$\\
~\textsc{EBOM} & \best{4.33} & \best{3.47} & 3.05 & 2.74 & 2.54 & 2.51 & 2.40 & 2.40 & 2.57\\
~\textsc{FSBNDM$qs$~} & 4.66$^{(2,0)}$ & 3.55$^{(3,1)}$ & 2.77$^{(3,1)}$ & 2.39$^{(3,1)}$ & 2.39$^{(3,1)}$ & 2.38$^{(3,1)}$ & 2.41$^{(3,1)}$ & 2.42$^{(3,1)}$ & 2.43$^{(3,1)}$\\
~\textsc{KBNDM} & 7.84 & 5.49 & 4.22 & 3.59 & 3.28 & 3.08 & 3.04 & 3.03 & 3.03\\
~\textsc{SBNDM$q$} & 5.75$^{(2)}$ & 4.18$^{(2)}$ & 3.13$^{(4)}$ & 2.43$^{(4)}$ & 2.52$^{(4)}$ & 2.50$^{(4)}$ & 2.52$^{(4)}$ & 2.51$^{(4)}$ & 2.52$^{(4)}$\\
&&&&&&&&&\\[-0.1cm]
\hline
&&&&&&&&&\\[-0.1cm]
~\textsc{FS-$w$} & 6.05$^{(6)}$ & 4.25$^{(6)}$ & 3.39$^{(6)}$ & 2.89$^{(6)}$ & 2.73$^{(6)}$ & 2.54$^{(6)}$ & 2.43$^{(6)}$ & 2.40$^{(6)}$ & 2.39$^{(6)}$\\
~\textsc{FJS} & 7.06 & 25.33 & 3.68 & 2.95 & 2.96 & 2.81 & 3.18 & 3.42 & 3.83\\
~\textsc{HASH$q$} & 19.96$^{(3)}$ & 8.34$^{(3)}$  & 5.02$^{(3)}$  & 3.68$^{(5)}$  & 3.17$^{(5)}$ & 2.95$^{(5)}$ & 2.96$^{(5)}$ & 2.76$^{(5)}$ & 2.65$^{(5)}$\\
~\textsc{IOM} & 9.37 & 6.67 & 5.26 & 4.38 & 3.96 & 3.73 & 3.47 & 3.30 & 3.20\\
~\textsc{WOM} & 9.98 & 7.01 & 5.28 & 4.32 & 3.91 & 3.53 & 3.25 & 3.11 & 3.02\\
&&&&&&&&&\\[-0.1cm]
\hline
&&&&&&&&&\\[-0.1cm]
~\textsc{WFR}   & 8.25 & 6.47 & 4.67 & 3.61 & 2.78 & 2.47 & 2.17 & 1.89 & 1.75\\
~\textsc{WFR$q$} & 5.20$^{(4)}$ & 3.89$^{(4)}$ & 3.08$^{(4)}$ & 2.42$^{(4)}$ & \best{2.08$^{(4)}$} & \best{1.97$^{(4)}$} & \best{1.91$^{(4)}$} & \best{1.76$^{(4)}$} & \best{1.69$^{(4)}$}\\
~\emph{speed-up} & +20\% & +12\% & +7.6\% & +2.9\% & -5.4\% & -5.7\% & -7.72\% & -12\% & -15\%\\
&&&&&&&&&\\[-0.1cm]
\hline
&&&&&&&&&\\[-0.1cm]
~\textsc{EPSM} & 6.72 & 6.36 & 2.86 & 2.13 & 1.94 & 1.94 & 1.92 & 1.86 & 1.87\\
~\textsc{TSO$q$} & 5.54$^{(5)}$ & 4.05$^{(5)}$ & 3.26$^{(5)}$ & 2.61$^{(5)}$ & 2.23$^{(5)}$ & - & - & - & -\\
&&&&&&&&&\\
\hline
\end{tabular*}
\end{scriptsize}
\end{center}
\caption{\label{exp_tab_english}Experimental results on a natural language sequence.}
\end{table}

Experimental results show that the BSDM$q$ algorithm obtains the best running times among previous solutions, especially in the case of long patterns. However it is second to the EBOM algorithm in the case of short patterns.

Our proposed \textsc{Wfr} algorithm performs well in several cases and turns out to be competitive against previous solutions. It even turns out to be faster than the BSDM$q$ algorithm in the case of very long patterns ($m \geq 256$), since the shift performed by the \textsc{Wfr} algorithm are longer on average than the shifts performed by the BSDM$q$ algorithm.

When the \textsc{Wfr} algorithm is implemented using unchained-loop, the performance increases further. Specifically, the \textsc{Wfr}$q$ algorithm turns out to be the fastest solution for patterns with a moderate length and for long patterns ($m\geq 32$). Better performances are obtained in the case of small alphabets, where the gain is up to 25\%, whereas in the case of large alphabets the gain is up to 14\%.

\section{Conclusions} \label{sec:conclusions}
In this paper we investigated a weak-factor-recognition approach to the exact string matching problem and devised an algorithm which, despite its quadratic worst case time complexity, shows a sublinear behaviour in practical cases.
Experimental results show that under suitable conditions, our algorithm obtains better running times than the most efficient algorithms known in literature.
It would be interesting to investigate whether multiple hashing functions can be used to reduce the number of false positives in the searching phase, in order to obtain better results.
A deeper analysis of the implemented hash function and of the implemented data structure will be performed in future works.


\section*{Acknowledgments}
This work has been supported by G.N.C.S., Istituto Nazionale di Alta Matematica ``Francesco Severi''.

\bibliographystyle{plain}

\begin{thebibliography}{10}

\bibitem{ACR99}
C. Allauzen, M. Crochemore, M. Raffinot.
\newblock Factor oracle: a new structure for pattern matching.
\newblock {\em in SOFSEM'99, Lecture Notes in Computer Science}, Vol.\ 1725, pages 291--306, 1999.

\bibitem{BYG92}
R.~Baeza-Yates and G.~H. Gonnet.
\newblock A new approach to text searching.
\newblock {\em Commun. ACM}, 35(10):74--82, 1992.

\bibitem{CF05}
D. Cantone and S. Faro.
\newblock Fast-Search Algorithms: New Efficient Variants of the Boyer-Moore Pattern-Matching Algorithm.
\newblock {\em Journal of Automata, Languages and Combinatorics}, 10(5/6):589--608, 2005.

\bibitem{CF14}
D. Cantone and S. Faro.
\newblock Improved and Self-Tuned Occurrence Heuristics.
\newblock {\em Journal of Discrete Algorithms}, 28:73--84, 2014.

\bibitem{CFG12}
D. Cantone, S. Faro, and E. Giaquinta.
\newblock A compact representation of nondeterministic (suffix) automata for
  the bit-parallel approach.
\newblock {\em Inf. Comput.}, 213:3--12, 2012.

\bibitem{charras04}
C.~Charras and T.~Lecroq.
\newblock {\em Handbook of exact string matching algorithms}.
\newblock King's College, 2004.



\bibitem{crochemore94}
M.~Crochemore, A.~Czumaj, L.~Gasieniec, S.~Jarominek, T.~Lecroq, W.~Plandowski,
  and W.~Rytter.
\newblock Speeding up two string-matching algorithms.
\newblock {\em Algorithmica}, 12(4):247--267, 1994.

\bibitem{DPST10}
B.~Durian, H.~Peltola, L.~Salmela, and J.~Tarhio.
\newblock Bit-parallel search algorithms for long patterns.
\newblock In {\em SEA, Lecture Notes in Computer Science}, vol.\  6049, pages 129--140, 2010.

\bibitem{Durian14}
B. Durian, T. Chhabra, S.S. Ghuman, T. Hirvola, H. Peltola, J. Tarhio.
\newblock Improved Two-Way Bit-parallel Search. 
\newblock In {\em Proc.\ of Stringology}, pages 71--83, 2014.

\bibitem{FK14}
S Faro and O. K\"ulekci.
\newblock Fast and Flexible Packed String Matching.
\newblock {\em Journal of Discrete Algorithms}, 28:61--72, 2014.

\bibitem{FL09}
S. Faro and T. Lecroq.
\newblock Efficient Variants of the Backward-Oracle-Matching Algorithm. 
\newblock {\em Int. J. Found. Comput. Sci.} 20(6):967--984, 2009.

\bibitem{FTBDM16}
S. Faro, T. Lecroq, S. Borz\`i, S. Di Mauro,  A. Maggio.
\newblock The String Matching Algorithms Research Tool. 
\newblock In {\em Proc.\ of Stringology}, pages 99--111, 2016.
               
\bibitem{FL10}
S. Faro and T. Lecroq.
\newblock The exact string matching problem: a comprehensive experimental
  evaluation.
\newblock {\em CoRR}, abs/1012.2547, 2010.


\bibitem{FL12}
S. Faro and T. Lecroq.
\newblock A Fast Suffix Automata Based Algorithm for Exact Online String Matching.
\newblock In {\em CIAA, Lecture Notes in Computer Science}, vol.\  7381, pages 149--158, 2012.

\bibitem{FL12b}
S. Faro and T. Lecroq.
\newblock A Multiple Sliding Windows Approach to Speed Up String Matching Algorithms.
\newblock In {\em SEA, Lecture Notes in Computer Science}, vol.\ 7276, pages 172--183, 2012.

\bibitem{FL13}
S. Faro and T. Lecroq.
\newblock The exact online string matching problem: a review of the most recent
  results.
\newblock {\em ACM Computing Surveys}, 45(2): Article No.\ 13, 2013.

\bibitem{FG05}
K. Fredriksson and S. Grabowski. 
\newblock Practical and Optimal String Matching. 
\newblock {\em SPIRE, Lecture Notes in Computer Science}, vol.\ 3772, pages 376--387, 2005.


\bibitem{KMP77}
D.~E. Knuth, J.~H. Morris, Jr, and V.~R. Pratt.
\newblock Fast pattern matching in strings.
\newblock {\em SIAM J. Comput.}, 6(1):323--350, 1977.

\bibitem{Lec07}
T.~Lecroq.
\newblock Fast exact string matching algorithms.
\newblock {\em Inf. Process. Lett.}, 102(6):229--235, 2007.

\bibitem{NR98}
G.~Navarro and M.~Raffinot.
\newblock A bit-parallel approach to suffix automata: Fast extended string
  matching.
\newblock In {\em CPM, Lecture Notes in Computer Science}, vol.\ 1448, pages 14--33, 1998.


\bibitem{PT11}
H. Peltola, J. Tarhio.
Variations of Forward-SBNDM. 
\newblock In {\em Proc.\ of Stringology}, pages 3--14, 2011.

\bibitem{yao79}
A.~C. Yao.
\newblock The complexity of pattern matching for a random string.
\newblock {\em SIAM J. Comput.}, 8(3):368--387, 1979.

\end{thebibliography}

\end{document}